\documentclass[%
reprint,
showkeys,
nofootinbib,
amsmath,amssymb,
aps,
prd,
]{revtex4-2}
\usepackage[marginal]{footmisc}
\usepackage{graphicx}
\usepackage{dcolumn}
\usepackage{bm}
\usepackage[colorlinks,allcolors=black,CJKbookmarks=True]{hyperref}
\usepackage{subcaption} 

\begin{document}
	
\title{Critical behavior and ultraviolet scaling of induced gravitational waves from an early matter-dominated era}

\author{Zhen-Min Zeng$^{1,2}$ }
\email{cengzhenmin@itp.ac.cn}

\author{Cheng-Jun Fang$^{1,2}$}
\email{fangchengjun@itp.ac.cn}

\author{Zong-Kuan Guo$^{1,2,3}$}
\email{guozk@itp.ac.cn}

\affiliation{$^{1}$Institute of Theoretical Physics, Chinese Academy of Sciences, Beijing 100190, China }
\affiliation{$^{2}$School of Physical Sciences, University of Chinese Academy of Sciences, Beijing 100049, China }
\affiliation{$^{3}$School of Fundamental Physics and Mathematical Sciences, Hangzhou Institute for Advanced Study, University of Chinese Academy of Sciences, Hangzhou 310024, China}

\begin{abstract}
Critical behavior and ultraviolet scaling of induced gravitational waves (GWs) from an early matter-dominated (eMD) era are studied in the context of primordial black hole evaporation. The depth of the eMD is characterized by the minimum parameter of the equation of state $\omega_{\min}$ that the Universe can attain during this phase. We identify a critical value $\omega_{c}\sim 7.3\times10^{-3}$ that separates two regimes. For $\omega_{\min}<\omega_{c}$, the GW peak lies at the non-linear cut-off point and requires non-linear dynamics. For $\omega_{\min}>\omega_{c}$, the peak originates from modes that reenter near the matter-radiation equality, and the ultraviolet tail follows a distinct scaling $k^{-3/2}$. This critical behavior provides a clear definition of deep versus shallow eMD and a robust spectral signature for future GW observations.
\end{abstract}
\maketitle

\section{Introduction}
Gravitational waves (GWs) offer a powerful window into the early Universe, particularly the epoch preceding big bang nucleosynthesis (BBN). They can be generated by a variety of mechanisms, including inflation~\cite{Guth:1980zm,Starobinsky:1980te,Mukhanov:1981xt,guzzetti_gravitational_2016,riotto_inflation_2017}, first-order phase transitions~\cite{Witten:1984rs,Kosowsky:1991ua,Hindmarsh_2021,Mazumdar_2019}, preheating~\cite{Kofman_1997,Greene:1997fu,Felder:2001kt,Dufaux_2007}, and topological defects~\cite{VILENKIN1985263,Hindmarsh_1995,Saikawa_2017}. Among these, GWs induced by first-order scalar perturbations have attracted significant attention~\cite{Ananda_2007,Baumann_2007,Bartolo_2007,Mangilli_2008,Saito_2009,assadullahi_gravitational_2009,Assadullahi_2010,Kawasaki_2013,Kohri_2018,Dom_nech_2020}. Their production is highly sensitive to the Universe’s thermal history, as they arise from first-order scalar perturbations during horizon reentry. The efficiency of this process depends strongly on the background evolution: deviations from radiation domination, such as an early matter-dominated (eMD) phase, can alter the dynamics of perturbations and the expansion rate, thereby modifying the amplitude and spectral shape of the induced GWs.

Although the standard cosmological model assumes radiation domination(RD) before BBN, this period remains poorly constrained by direct observations. Non-standard phases, such as eMD, are viable alternatives and can naturally emerge in scenarios involving a decaying dark sector~\cite{Zhang_2015,hook_causal_2021,Erickcek_2021,Erickcek_2022,Nelson_2018,Blinov_2020}, primordial black holes (PBHs)~\cite{Sasaki:2018dmp,inomata_enhancement_2019,inomata_gravitational_2019,carr_constraints_2021,escriva_primordial_2023,escriva_primordial_2024,Dom_nech_2022}, post-inflationary reheating~\cite{Jedamzik_2010,Erickcek_2011,Fan_2014}, or solitonic objects like Q balls~\cite{kusenko_supersymmetric_1998,Kasuya_2023}.

Previous studies have explored how a deep eMD era before BBN would modify the evolution of induced GWs, leading to a characteristic enhancement in their present-day spectrum~\cite{inomata_gravitational_2020,inomata_enhancement_2019,inomata_gravitational_2019,pearce_gravitational_2024,kumar_towards_2024,pearce_using_2025}. In general, this enhancement requires a fast transition rate from the eMD era to the RD era. 
During the eMD epoch, the gravitational potential does not decay, even on subhorizon scales. The amplification of induced GWs comes from the fast oscillation of the scalar potential after the transition, especially the mode entering the horizon deep in the eMD era. The earlier the mode enters the horizon, the faster the scalar potential oscillates in the RD era, and the stronger the induced GWs. However, in the eMD era, the density contrast of matter grows with time and may exceed the linear region before the Universe goes back to the RD era. Inevitably, there will be a cutoff scale $k_{cut}$ that keeps the linear perturbation theory valid.

Unfortunately, the presence of a cutoff scale limits both the predictability and detectability of GWs generated by this mechanism. Previous studies, which neglected contributions from nonlinear modes, concluded that for a prolonged eMD epoch, the GW spectrum always peaks at the cutoff scale. However, the evolution in the non-linear regime beyond $k > k_{\rm cut}$ remains unknown, and the  ultraviolet(UV) tail of the spectrum is poorly understood. This uncertainty makes it difficult to distinguish this mechanism from alternative sources of GWs. As a result, predictions of the GW spectrum near the cutoff scale are not robust and should only be regarded conservative estimates. 

In particular, if the eMD epoch lasts for a long duration, a reliable prediction of the GW spectrum---especially its behavior near the peak frequency---requires a proper treatment of non-linear dynamics. In contrast, when the eMD epoch is short, the peak of the GW spectrum may not depend on non-linear dynamics, since the most amplified mode is not necessarily the cutoff mode. In this case, the cutoff mode can even reenter the horizon during the eRD era, experiencing suppression before the onset of eMD. This implies that for a short-lived eMD epoch, the UV behavior of the spectrum may be investigated without detailed knowledge of the nonlinear regime.


In this paper, we revisit the mechanism of GW amplification that arises from a rapid transition between the eMD and RD eras. Our goal is to determine the critical depth of the eMD and, in particular, the dependence of the GW peak on the depth of this phase. Among the possible eMD realizations, the case of PBH evaporation is especially promising, as it provides the fastest transition from eMD to RD~\cite{pearce_using_2025}. We therefore focus on the PBH evaporation scenario, assuming a monochromatic PBH mass function. This makes our setup the most favorable for detection, since a broader mass distribution would suppress the GW amplitude. Under these assumptions, our analysis centers on the impact of the eMD depth on the GW spectrum, with the derived critical depth representing a conservative estimate for detectability.  Addressing these questions requires full numerical techniques to compute the GW spectrum without relying on the sudden-transition approximation, as the Universe may undergo only a very brief eMD phase.

We find that the eMD era exhibits a critical depth that separates two qualitatively different regimes of the induced GW spectrum. In the deep regime, the spectral peak lies at the nonlinear cutoff and its prediction requires nonlinear dynamics beyond linear theory. In the shallow regime, the peak instead arises from modes reentering near the matter-radiation equality, and the UV tail follows an approximate $k^{-3/2}$ scaling. This scaling, which has not been identified in previous studies, provides a distinctive spectral feature of shallow eMD. The existence of a critical depth thus offers a clear criterion for distinguishing deep and shallow eMD phases, and highlights the potential of induced GWs to serve as a sensitive probe of the early-Universe thermal history in future observations.

\section{Depth of the PBH dominated era}
In this work, we consider the case where there is an eRD epoch followed by an eMD epoch before nucleosynthesis. The eMD era is motivated by the PBHs that originated from inflation when large perturbations reenter the Horizon\cite{Ballesteros:2017fsr,Martin:2012pe,escriva_primordial_2023}. Once the PBHs form, they stay stable at an early time and will eventually evaporate through Hawking radiation. Their mass determines the evaporation time and their abundance determines the duration or depth of the eMD era. After the PBHs evaporate to radiation, the Universe follows the standard evolution in the RD era.

We denote the two times of the matter-radiation equality by $\tau_{eq1}$ and $\tau_{eq2}$. We mainly consider PBHs as the matter component, where their decay becomes extremely rapid around $\tau_{eq2}$. Hence, we identify $\tau_{eq2}$ with the onset of the RD era, denoted as $\tau_{eva}$, and will not distinguish between them throughout this paper.

\subsection{PBH dominated condition}
The initial mass of PBHs is some ratio of the Hubble volumn at the formation time~\cite{Carr:1974nx}:
\begin{equation}
    M_{PBH,i}\simeq \left.\gamma\frac{4}{3}\pi H^{-3}\rho\right|_{t=t_i},
\end{equation}
where $\gamma\sim0.2$ is the fraction of the PBH mass in the horizon mass at the formation~\cite{1975ApJ...201....1C}.
$\rho$ is the energy density and $H$ is the Hubble parameter. In the RD era, $H=1/2t$ and $\rho=3H^2M_{pl}^2$. So, the formation time of PBHs can be expressed as
\begin{equation}
    t_i\simeq2.24562\times10^{-30}s\left(\frac{M_{PBH,i}}{10^8 g}\right).
\end{equation}
PBH evaporation follows the equation
\begin{equation}
\label{eqn:hawking}
    \frac{d M_{PBH}}{dt}=-\frac{A}{M_{PBH}^2},
\end{equation}
 where $A=3.8\pi g_{H}(T_{PBH})M_{pl}^4/480$,and $g_{H}(T_{PBH})$ is the spin-weighted degree of freedom of the particles emitted from Hawking radiation with $T_{PBH}$. For PBHs with an initial mass less than $10^{11}g$, $g_{H}(T_{PBH})\approx108$~\cite{dolgov2000gravitationalwavesbaryogenesisdark}. Integrate Eq.~\eqref{eqn:hawking} from formation time to evaporation time
\begin{equation}
    \int_{M_{PBH,i}}^0 M_{PBH,i}^2dM_{PBH,i}=-A\int_{t_i}^{t_{eva}}dt.
\end{equation}
We obtain the evaporation time of PBH with mass $M_{PBH,i}$
\begin{equation}
    t_{eva}\simeq 4\times10^{-4}s\left(\frac{M_{PBH,i}}{10^8 g}\right)^3.
\end{equation}
In general, the decay rate can be expressed as
\begin{equation}
    \Gamma\equiv-\frac{1}{M_{PBH}}\frac{dM_{PBH}}{dt}=\frac{1}{3(t_{eva}-t)}.
    \label{eqn:ga}
\end{equation}
Once PBHs forms, if they stay stable(far before the evaporation time $t_{eva}$), they redshift slower than radiation, and their energy density contrast between PBHs and radiation is growing, which can be characterized by 
\begin{equation}
    \beta(t)\equiv\frac{\rho_{PBH}}{\rho_r}\approx\frac{\rho_{PBH,i}}{\rho_{r,i}}\frac{a(t)}{a_i}=\beta_i\frac{a(t)}{a_i},
\end{equation}
where $\rho_{PBH,i}$ and $\rho_{r,i}$ are the initial energy density of PBH and radiation. Here we ignore the Hawking radiation of PBHs using the relation $\rho_{PBH}(t)a^{3}(t)=\rho_{PBH,i}a^{3}_i$ and $\rho_{r}(t)a^4(t)=\rho_{r,i}a^4_i$. PBH will dominate the Universe if $\beta_i>a(\tau_{eq1})/a_i$. 

All in all, whether there will be an eMD era or not is determined by PBHs' initial mass and the abundance, which originate from inflation, where the mass is determined by the specific structure of inflation potential, and their abundance is determined by the probability distribution function of the relevant curvature perturbation. In this paper, we specifically focus on the effect of depth of the eMD era on the GW spectrum.

\subsection{Nonlinear cutoff}

During the eMD epoch, the gravitational potential does not decay, even on subhorizon scales. However, the density contrast grows in time on all scales and is given by the Poisson equation:
\begin{equation}
    \delta_m=\frac{2k^2}{3\mathcal{H}^2}\phi,
\end{equation}
where $\delta_m\equiv\delta\rho_m/\rho_m$ and $\mathcal{H}$ is the comoving Hubble parameter. If the eMD epoch lasts long, there will be a critical mode which reenter the horizion deep in eMD era and its density contrast grows to $\delta_m\sim1$. This can be estimated by~\cite{Assadullahi_2010}
\begin{equation}
    k_{NL}\sim\sqrt{\frac{5}{2}}\mathcal{P}_{\mathcal{R}}^{-1/4}\mathcal{H}(\tau_r)\sim\frac{470}{\tau_r},
\end{equation}
 where $\mathcal{P}_{\mathcal{R}}\sim2.1\times 10^{-9}$ is the amplitude of the primordial curvature perturbation power spectrum obtained by CMB observation~\cite{plank2020}. The mode larger than $k_{NL}$ reaches the non-linear region and decouples from the Hubble flow.

However, if the eMD phase is short, this estimation breaks down since the critical cutoff mode may enter the horizon in the eRD era and at that time the matter is a subdominant component. So we cannot use the normal trick to determine the evolution of matter perturbations from the Einstein equations of gravitational potential. Instead, we imply the Mészáros equation, which is derived by the continuous equation and Euler equation~\cite{Baumann:2022mni}
\begin{equation}
    \frac{d^2\delta_m}{d s^2}+\frac{2+3s}{2s(1+s)}\frac{d\delta_m}{d s}-\frac{3}{2s(1+s)}\delta_m=0,
\end{equation}
where $s\equiv a/a_{eq}$ and the solutions are
\begin{equation}
    \delta_m\propto\left\{ 
    \begin{aligned}
        &1+\frac{3}{2}s,\\
        &-3\sqrt{1+s}+(1+\frac{3}{2}s)\ln{(\frac{\sqrt{1+s}+1}{\sqrt{1+s}-1})}.
    \end{aligned}
    \right.
\end{equation}
During the MD era, corresponding to the limit $s \gg 1$, the density contrast evolves as $\delta_m \propto a$, while in the RD era ($s \ll 1$), it grows only logarithmically, $\delta_m \propto \ln a$. This indicates that the growth of $\delta_m$ is much slower in the RD era than in the MD era. Consequently, if the duration of the eMD epoch is short, $k_{NL}\tau_r$ becomes much larger than 470, which has important implications for the prediction of GW. Previous studies have shown that for a long eMD period, the GW spectrum peaks in the maximal mode $k_{NL}$. However, for a shorter eMD epoch, the cutoff scale can be significantly larger, and some modes may reenter the horizon during the eRD era. These modes experience additional suppression of $a^{-2}$ before entering the eMD phase, which implies that the GW spectrum does not peak at $k_{NL}$. Instead, the dominant contribution comes from modes that reenter the horizon near the matter-radiation equality. This allows us to probe the UV behavior of the spectrum in the range $k_{eq1}<k<k_{NL}$. In such cases, the cutoff scale cannot be simply approximated as $k_{NL}\sim 470/\tau_{eva}$; rather, one needs to numerically solve the full evolution of perturbations. In the following, we do not distinguish $k_{NL}$ and $k_{cut}$.



\section{Evolution of background field and perturbations}
 In order to solve the background in conformal time, we rewrite the decay rate~[\ref{eqn:ga}] as differential equations. Background evolution can be expressed as
 \begin{eqnarray}
     \mathcal{H}^2&=&a^2(\rho_r+\rho_m)/3,\label{eqn:fri}\\
     \rho_m^\prime&=&-3\mathcal{H}\rho_m-\tilde{\Gamma}\rho_m,\label{eqn:rhom}\\
     \rho_r^\prime&=&-4\mathcal{H}\rho_r+\tilde{\Gamma}\rho_m,\label{eqn:rhor}\\
     \tilde{\Gamma}^\prime&=&\mathcal{H}\tilde{\Gamma}+3\tilde{\Gamma}^2,\label{eqn:gat}
 \end{eqnarray}
where $\tilde{\Gamma}\equiv a\Gamma$ is the conformal decay rate,, $\rho_m$ and $\rho_r$ are energy density of matter and background radiation.

We start our numerical calculation at the formation time of PBHs $\tau_i$ in the eRD era, 
where the abundance of PBHs is characterized by $\beta_i=\rho_m(\tau_i)/\rho_r(\tau_i)$. We assume $\beta_i\ll1$ throughout this paper. We can ignore the evaporation of PBHs at initial time and the scale factor can be expressed as
\begin{equation}
    \frac{a(\tau)}{a_i}=\frac{1}{\beta_i}\left[(2+\beta_i-2\sqrt{1+\beta_i})\left(\frac{\tau}{\tau_i}\right)^2+2(-1+\sqrt{1+\beta_i})\frac{\tau}{\tau_i}\right].
\end{equation}
We rescale the conformal time and scale factor at $\tau_i$, where $\tau/\tau_i\rightarrow\tilde{\tau}, a(\tau)/a_i\rightarrow\tilde{a}(\tilde{\tau})$. The energy density can be dimensionless as $\rho_m\tau_i^2/M_{pl}^2\rightarrow\tilde{\rho}_m,\rho_r\tau_i^2/M_{pl}^2\rightarrow\tilde{\rho}_r$. From now on, we omit the symbol ``$\tilde{\quad}$", just keep in mind that all variables are dimensionless and rescaled at $\tau_i$. Then the initial conditions can be expressed as
\begin{equation}
\begin{split}
     &a(\tau_i)=1, \quad a'(\tau_i)=\frac{2(1+\beta_i-\sqrt{1+\beta_i})}
     {\beta_i},\\&\rho_r(\tau_i)=\frac{27\beta_i^2}{4(-1+\beta_i+\sqrt{1+\beta_i})^2},\quad \rho_m(\tau_i)=\beta_i\rho_r(\tau_i).
\end{split}
\label{eqn:initial_bkg}
\end{equation}
 The initial condition for $\tilde{\Gamma}$ is more tricky. At formation time, $t_i\ll t_{eva}$ and $\Gamma(t_i)\approx1/(3t_{eva})$, which is determined by the initial mass of PBHs. 
 However, $t_{eva}$ cannot be determined since we need to integrate the scale factor in conformal time, which is unclear before solving the background equations. In practice, we can pre-set the numerical quantity of $t_{eva}$, then solve the background equations until the end of the transition $\tau_{eva}$, where $\rho_m(\tau_{eva})\ll\rho_r(\tau_{eva})$ and $\omega(\tau_{eva})\approx1/3$. Then the actual value of $t_{eva}$ can be determined by integrating the conformal time from $0$ to $\tau_{eva}$. Until now, we know exactly the initial mass of PBHs. 
 In general, only two parameters $\beta_i$ and $M_{PBH,i}$ can determine whether there will be the eMD era or not, as well as the depth of the eMD era, if any.

We then calculate perturbations in the conformal Newtonian gauge~\cite{ma_cosmological_1995,inomata_gravitational_2019}
\begin{eqnarray}
    \delta_m^\prime&=&-\theta_m+3\phi^\prime-\tilde{\Gamma}\phi-3\tilde{\Gamma}^2\frac{\theta_m}{k^2},\label{eqn:delatm}\\
    \theta_m^\prime&=&-\mathcal{H}\theta_m+k^2\phi,\label{eqn:thetam}\\
    \delta_r^\prime&=&-\frac{3}{4}(\theta_r-3\phi^\prime)+\tilde{\Gamma}\frac{\rho_m}{\rho_r}(\delta_m-\delta_r+\phi)+3\tilde{\Gamma}^2\frac{\rho_m}{\rho_r}\frac{\theta_m}{k^2},\label{eqn:delatr}\\
    \theta_r^\prime&=&\frac{k^2}{4}\delta_r+k^2\phi-\tilde{\Gamma}\frac{3\rho_m}{4\rho_r}(\frac{4}{3}\theta_r-\theta_m),\label{eqn:thetar}\\
    \phi^\prime&=&-\frac{k^2\phi}{3\mathcal{H}}-\mathcal{H}\phi-\frac{\mathcal{H}}{2}(\frac{\rho_m}{\rho_{tot}}\delta_m+\frac{\rho_r}{\rho_{tot}}\delta_r),\label{eqn:phinum}
\end{eqnarray}
where we neglect anisotropic stress, so $\phi$ is the only scalar perturbation of the metric. 
Here $\rho_{tot}=\rho_m+\rho_r$ is the total energy density. Solving the above equations in superhorion limit~($\tilde{\Gamma}=0$), 
we get the initial condition in the eRD era.
\begin{equation}
    \delta_{m,i}=-\frac{3}{2}\phi_i,\quad\delta_{r,i}=-2\phi_i,\quad\theta_{m,i}=\theta_{r,i}=\frac{k^2\tau}{2}\phi_i,
\end{equation}
where $\phi_i$ is the primordial perturbation generated from inflation. It is worth mentioning that our choice is the adiabatic initial conditions. We solve the above perturbation equations until $\tau_{eva}$ since $\tilde{\Gamma}$ will be singular when approaching $\tau_{eva}$. To solve the evolution of perturbations after $\tau_{eva}$, we use $\delta_{r}(\tau_{eva})$, $\theta_{r}(\tau_{eva})$, and $\phi(\tau_{eva})$ as the initial condition of the following equations:
\begin{equation}
    \begin{split}
        \delta_r^\prime&=-\frac{3}{4}(\theta_r-3\phi^\prime)\\
    \theta_r^\prime&=\frac{k^2}{4}\delta_r+k^2\phi\\
    \phi^\prime&=-\frac{k^2\phi}{3\mathcal{H}}-\mathcal{H}\phi-\frac{\mathcal{H}}{2}\delta_r,
    \end{split}
\end{equation}
Once $\beta_i$ and $t_{\rm eva}$ are specified, that is, the initial abundance and mass of PBHs are fixed, the subsequent evolution of both the background and perturbations is fully determined. Next, we present the numerical method to calculate the GW spectrum.

\section{Numerical method to calculate the induced GW spectrum}
Given that both the background and perturbations are computed numerically, 
induced GWs must also be derived through numerical integration.
The energy density of induced GWs is given by
\begin{equation}
    \Omega_{GW}(k,\tau_0)h^2=\Omega_{r,0}h^2\frac{1}{24}(\frac{k}{\mathcal{H}_c})^2\overline{\mathcal{P}_h(k,\tau_c)},
\end{equation}
where $\Omega_{r,0}h^2\approx4.2\times10^{-5}$ is the current radiation energy parameter, $\mathcal{P}_h$ is the power spectrum of GWs. 
The subscript ``c" represents the time when the mode of interest becomes well inside the horizon and the energy density of GWs becomes almost constant. $\overline{\mathcal{P}}_h$ is the time average per period $T$ around $\tau$, explicitly in $\bar{x}(\tau)=\frac{1}{T}\int_{\tau-T}^{\tau}x(\tilde{\tau})d\tilde{\tau}$. The power spectrum can be calculated by
\begin{equation}
\begin{split}
    \mathcal{P}_{h}(k,\tau)& = \frac{4}{81a^2(\tau)}
\int_{\substack{|k_1 -k_2| \le k \le k_1 +k_2}}\hspace{-5.5em} d\ln{k_1} \, d\ln{k_2} I^2(k,k_1,k_2,\tau)\\
&\times\frac{(k_1^2-(k^2-k_2^2+k_1^2)^2/(4k^2))^2}{k_1k_2k^2}\mathcal{P}_{\mathcal{R}}(k_1)\mathcal{P}_{\mathcal{R}}(k_2),
\end{split}
\label{eqn:phk}
\end{equation}
where $\mathcal{P}_{\mathcal{R}}(k)=\Theta(k_{max}-k)A_{s}(k/k_*)^{n_s-1}$ is the power spectrum of primordial curvature perturbations with $A_s\approx2.1\times10^{-9}$ being the amplitude at the pivot scale, $n_{s}\approx0.97$ the spectral tilt and $k_*=0.05\mathrm{Mpc^{-1}}$ the pivot scale~\cite{plank2020}. $I(k,k_1,k_2,\tau)$ is the kernel function defined by
   \begin{equation}
       \begin{split}
           &I(k,k_1,k_2,\tau)=4k^2\int_0^\tau d\tilde{\tau}a(\tilde{\tau})G_k(\tau,\tilde{\tau})\Bigg[2T_{k_1}(\tilde{\tau})T_{k_2}(\tilde{\tau})\Bigg.\\
           &\Bigg.+\frac{4}{3(1+\omega(\tilde{\tau}))}\left(T_{k_1}(\tilde{\tau})+\frac{T_{k_1}^\prime(\tilde{\tau})}{\mathcal{H}(\tilde{\tau})}\right)\left(T_{k_2}(\tilde{\tau})+\frac{T_{k_2}^\prime(\tilde{\tau})}{\mathcal{H}(\tilde{\tau})}\right)\Bigg],
       \end{split}
       \label{eqn:kernal}
   \end{equation}
where $G_{k}(\tau,\Tilde{\tau})$ is the Green function for tensor perturbation. $T_{k}(\tau)$ is the transfer function of the scalar potential, where $\phi_{k}(\tau)=T_{k}(\tau)\phi_{i}$. $\omega(\tau)=\rho_r(\tau)/3(\rho_m(\tau)+\rho_r(\tau))$ is the equation of the state parameter.

We can combine two independent homogeneous solution to obtain the Green function
\begin{equation}
    G_{k}(\tau,\tilde{\tau})=\frac{v_{1k}(\tau)v_{2k}(\tilde{\tau})-v_{1k}(\tilde{\tau})v_{2k}(\tau)}{v_{1k}^\prime(\tilde{\tau})v_{2k}(\tilde{\tau})-v_{1k}(\tilde{\tau})v_{2k}^\prime(\tilde{\tau})}\Theta(\tau-\tilde{\tau}),
    \label{eqn:greenfun}
\end{equation}
where the two homogeneous solutions $v_{ik}=ah_{k}$ can be obtained by equation of motion of tensor perturbation
\begin{equation}
    \left(\partial_\tau^2+k^2-\frac{1-3\omega(\tau)}{2}\mathcal{H}^2 \right)v_{ik}=0,
\end{equation}

The average square of the kernel can be obtained as~\cite{abe_induced_2021,abe_translating_2023}
\begin{equation}
\begin{split}
     \overline{I^2(k,k_1,k_2,\tau)}\approx I_2^2(k,k_1,k_2,\tau)\overline{v_{1k}^2(\tau)}\\+I_1^2(k,k_1,k_2,\tau)\overline{v_{2k}^2(\tau)}\\
     -2I_1(k,k_1,k_2,\tau)I_2(k,k_1,k_2,\tau)\overline{v_{1k}(\tau)v_{2k}(\tau)},
\end{split}
\end{equation}
 where $I_n(k,k_1,k_2,\tau)$ is obtained by splitting the Green function Eq.~\eqref{eqn:greenfun} into
two contributions.

 We conclude the steps to calculate the GWs as follows.
 \begin{itemize}
     \item Use Eqs.~\eqref{eqn:fri}-\eqref{eqn:gat} to solve the background equations from $\tau_i$ to $\tau_{eva}$ with the initial condition Eq.~\eqref{eqn:initial_bkg}. 
         Continuous background derived analytically during the late RD epoch, valid for $\tau>\tau_{eva}$.
     \item Find the cut-off scale $k_{cut}$ by solving the perturbation evolution Eqs.~\eqref{eqn:delatm}-\eqref{eqn:phinum}, 
     where the cut-off scale is defined as $\delta_{m,k_{cut}}\sim1$ at the end of transition.
     \item Choose the range of mode $[k_{min},k_{max}]$ in which we are interested. $k_{min}\tau_{rh}=1/100$ is the mode that enters the horizon in the late RD era. $k_{max}=k_{cut}$ is the cutoff scale.
     \item For a given mode $k\in[k_{min},k_{max}]$, we obtain the numerical solution of $v_{1k},v_{2k}$ using the analytical solution as the initial condition when they are well outside the horizon during the eRD epoch.
     \item For each wave number $k$ computed in the previous step, we discretize the $k_{*}\in[10^{-1},10^1]k$ interval into 250 logarithmically spaced bins, numerically solve the perturbation equations Eqs.~\eqref{eqn:delatm}-\eqref{eqn:phinum} for each mode, and systematically archive the resultant data.
     \item Combining $k_1$ and $k_2$ in the integration domain defined in Eq.~\eqref{eqn:phk}, where we evaluate the associated kernel functions specified in Eq.~\eqref{eqn:kernal}. Numerical integration is implemented through a trapezoidal quadrature scheme with a bidimensional grid of approximately $250\times 250$ sampling points.
     \item Calculate the time average of the GW power spectrum.
 \end{itemize}
Following the above steps, we can numerically obtain the GW power spectrum. 
Note that momentum integration only needs modes with $k_{*}\in[10^{-1},10^1]k$, 
which is sufficient since the dominant contribution arises from
$|\vec{k_*}|+|\vec{k}-\vec{k_*}|\sim \sqrt{3}|\vec{k}|$. The GW spectrum that we obtained covers the range $[10k_{min},0.1k_{cut}]$.

\section{Numerical results}
\begin{figure}
     \includegraphics[width=1\linewidth]{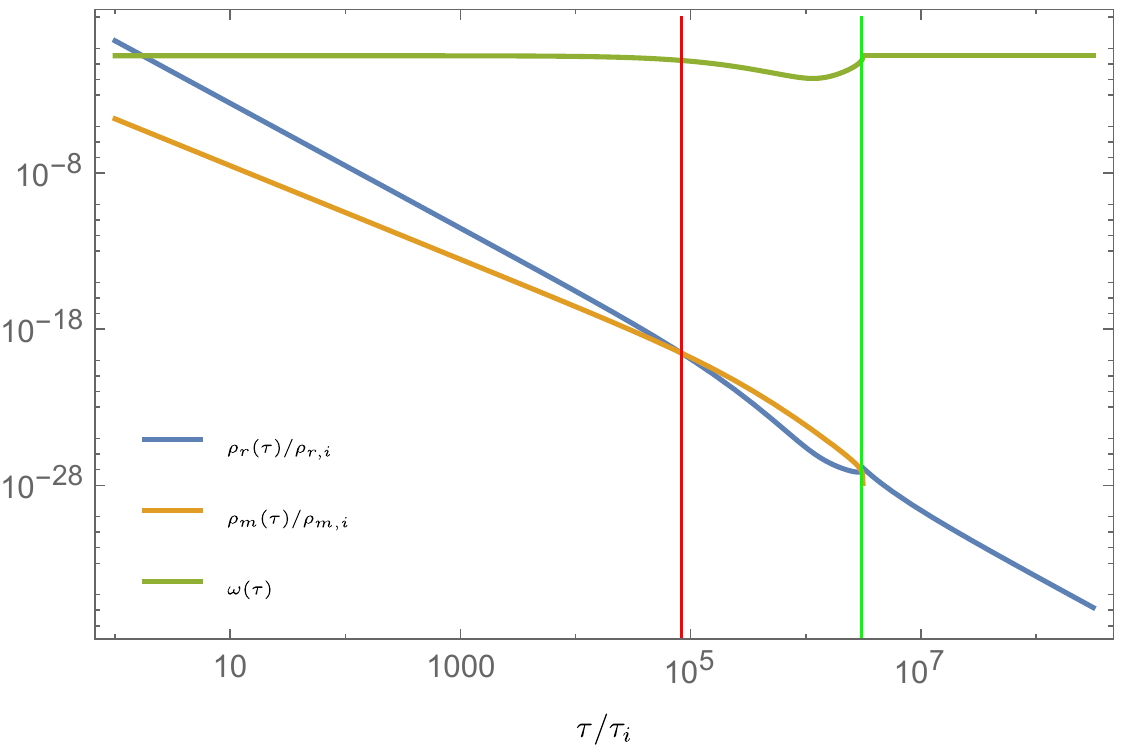}
     \captionsetup{justification=raggedright,singlelinecheck=false}
     \caption{Evolution of the equation of state and energy density of matter and radiation. The vertical lines represents former matter radiation equal time $\tau_{eq1}$~(red) and the latter matter radiation equal time $\tau_{eq2}$~(green) respectively.}
     \label{fig:backg}
 \end{figure}

\begin{figure}
     \includegraphics[width=1\linewidth]{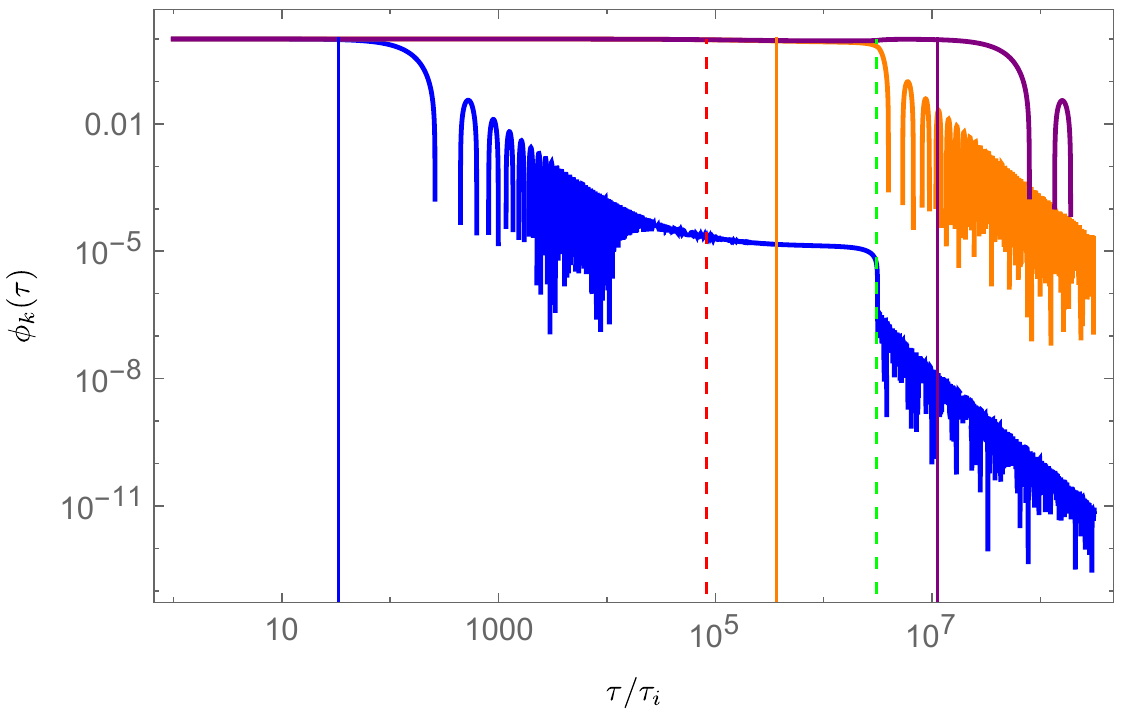}
     \captionsetup{justification=raggedright,singlelinecheck=false}
     \caption{Evolution of the scalar perturbations $\phi_{k}(\tau)$. Three curves colored blue, orange and purple represent the numerical solution of mode $k_{cut}=91700/\tau_{eq2}$(cutoff mode), $k=12/\tau_{eq2}$ and $k=0.3/\tau_{eq2}$, respectively. Vertical line labels the time of their reentry. 
     Red and Green dashed line labels $\tau_{eq1}$ and $\tau_{eq2}$. }
     \label{fig:phicom}
 \end{figure}

We start our calculation by setting $\beta_i=10^{-5}$, choosing different $t_{eva}$, i.e., different initial mass of PBH $M_{PBH,i}$ to determine different depths of the eMD era. For each pair of parameters, we solve the background and perturbations. In Fig.~\ref{fig:backg}, we show the evolution of the background. The smallest equation-of-state parameter
attainable during the eMD epoch is $\omega_{min}=1.18\times 10^{-2}$. 
The duration of the eMD epoch is short in this case. We also show the evolution of the corresponding perturbations in Fig.~\ref{fig:phicom}, for which we choose three typical modes: cutoff mode, modes that reenter the horizon during the eMD era, and modes that reenter the horizon in the late RD era. 
The dependence of $k_{cut}$. We can see that the cutoff mode reenters the Horizon in eRD era and experiences fast oscillation and decreases as the power of $a^{-2}$. At the evaporation time $t_{eva}$, it is suppressed by $\mathcal{O}(0.1)$, which originated from the finite transition time. We can see that the scalar perturbation of the mode reentering the Horizon in the MD era is much larger than the case of the cutoff mode, which implies that in this case the most enhanced mode in GW spectrum is not the cutoff mode.

\begin{figure}
    \centering
    \includegraphics[width=1\linewidth]{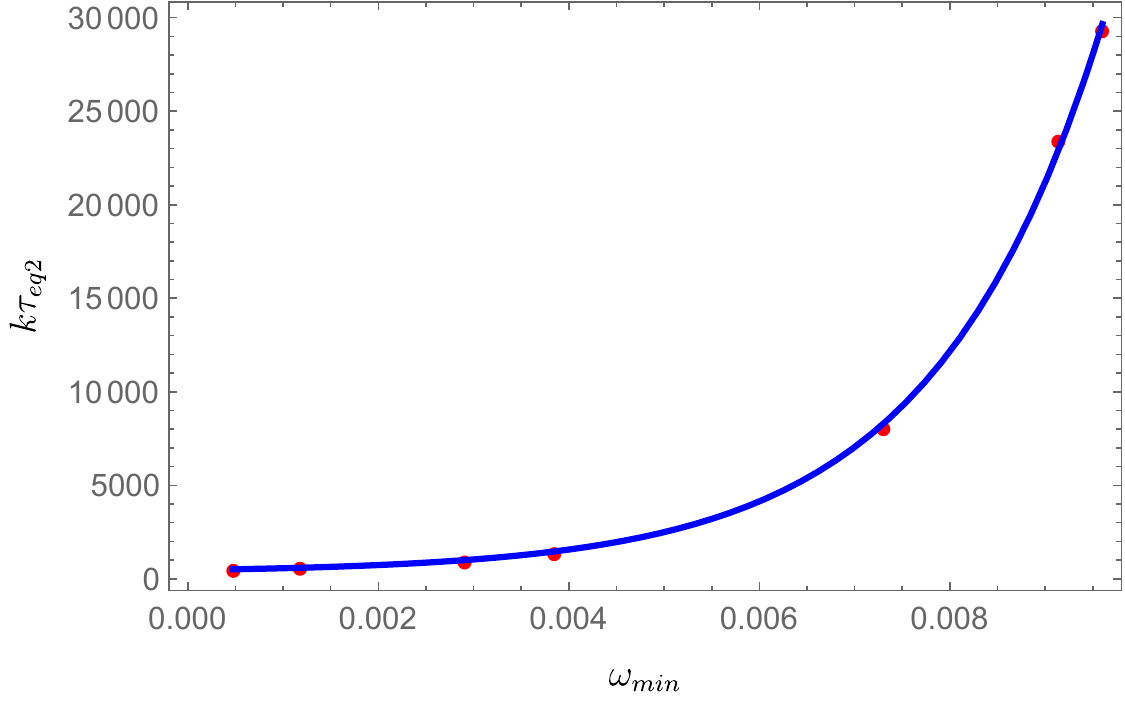}
    \captionsetup{justification=raggedright,singlelinecheck=false}
    \caption{The relation between the cutoff mode $k_{cut}$ and the depth of eMD epoch, characterized by $\omega_{min}$}
    \label{fig:ktw}
\end{figure}

We also present $k_{\rm cut}\tau_{eq2}$ as a function of $\omega_{\min}$. In the deep eMD limit, where $\omega_{\min}\to0$, we find $k_{cut}\tau_{eq2}\simeq470$. The dependence of $k_{cut}\tau_{eq2}$ on the depth of the eMD epoch is shown in Fig.~\ref{fig:ktw}. As the duration of the eMD decreases, $k_{cut}\tau_{eq2}$ increases approximately exponentially. The red dots represent our numerical results, while the blue line corresponds to the fitted function, which can be expressed as
\begin{equation}
    k_{cut}\tau_{eq2}=125.7e^{567.9\omega_{min}}+344.3
\end{equation}

We illustrate the dependence of the GW spectrum on the depth of the eMD epoch in Fig.~\ref{fig:phicom}. For long eMD durations, such as $\omega_{\min}=1.18\times10^{-3}$ and $\omega_{\min}=2.9\times10^{-3}$, the maximum frequency appears at $0.1k_{\rm cut}$, since to calculate this mode, we only integrate modes within $[0.01k_{\rm cut},k_{\rm cut}]$, thus excluding contributions from nonlinear dynamics. This ensures the reliability of our numerical results. As the eMD epoch becomes shorter, the spectral peak shifts from $0.1k_{\rm cut}$ to the mode that reenters the horizon slightly before $\tau_{eq1}$, which can be approximately expressed as $k_{\rm peak}\sim20/\tau_{eq1}$. As you can see in the gray dots (peak frequency) in Fig.~\ref{fig:phicom}, for short durations of eMD, they align nearly vertically. This behavior arises because we fix $\beta_i=10^{-5}$, while $M_{\rm PBH,i}$ has little effect on $\tau_{eq1}$. This is the balance of the resonance oscillation in the late RD era and the suppression in the eRD era.  In particular, we find that the UV tail scales as $k^{-3/2}$, where the gray dashed line shows this scaling. The critical case is labeled by the brown curve ($\omega_{c}=7.3\times10^{-3}$), where the peak mode coincides with the cutoff mode. If $\omega<\omega_{c}$, it is inevitable that the dynamics of the non-linear mode will be required to get the exact peak behavior of the GW spectrum. However, if $\omega>\omega_{c}$, the peak is located in the linear region and $k_{\rm peak}\sim20/\tau_{eq1}$.  This feature was not discussed in previous studies, which may provide valuable information for future investigations of GW signals from an early matter-dominated epoch.

  \begin{figure}
     \centering
     \includegraphics[width=1\linewidth]{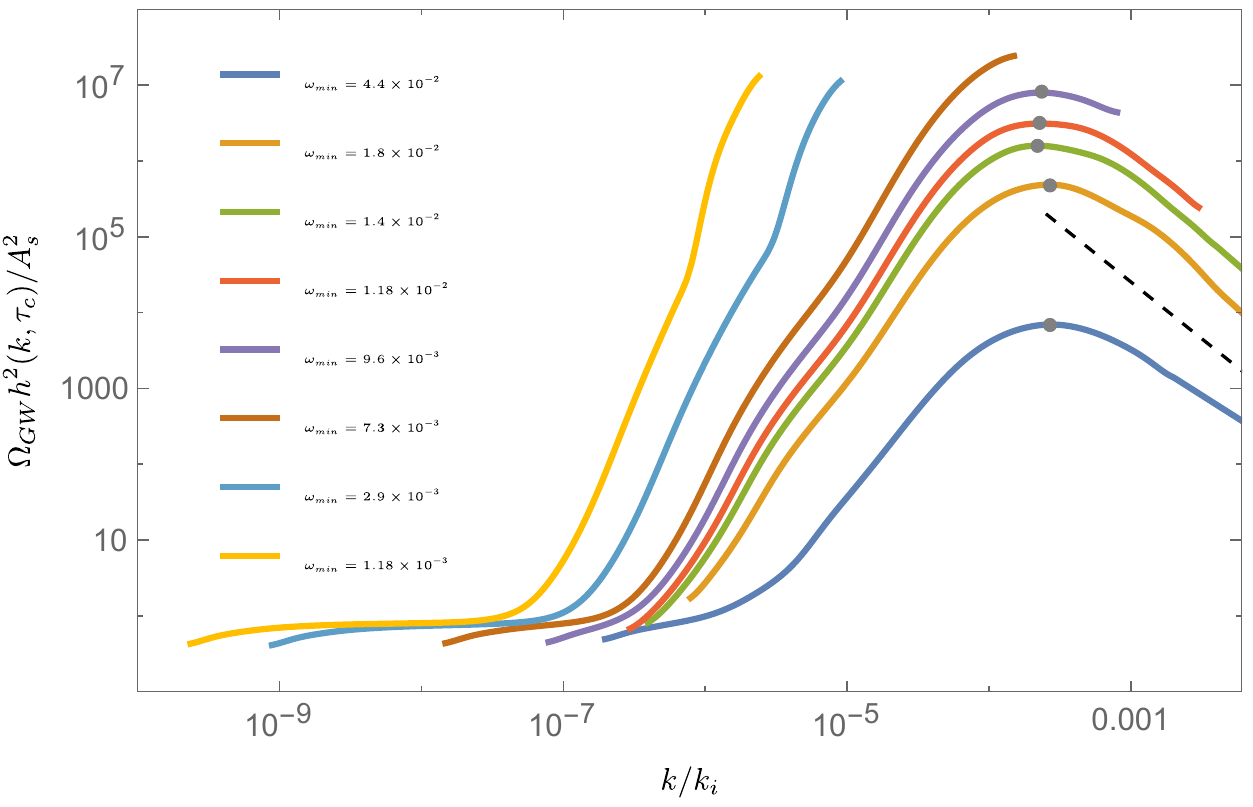}
     \caption{Energy spectrum of GW at the time $\tau_c$ where the mode of interest becomes well inside the horizon. }
     \label{fig:omega}
 \end{figure}

\section{Conclusions and discussions}

We have studied induced GWs from an eMD era caused by PBH evaporation. By numerically solving the background and perturbations without the sudden-transition approximation, we identified a critical depth at $\omega_{\min,c}\sim 7.3\times10^{-3}$ that separates two regimes. For $\omega<\omega_{\min,c}$, the GW spectrum peaks at the nonlinear cutoff and depends on nonlinear dynamics. For $\omega>\omega_{\min,c}$, the peak arises from modes that reenter near the matter-radiation equality, and the UV tail follows a distinct scaling $k^{-3/2}$.

This critical behavior provides a robust criterion for distinguishing deep and shallow eMD phases. The $k^{-3/2}$ scaling offers a clear spectral feature absent in other GW production mechanisms, while amplification at the critical depth may bring the signal within reach of next-generation detectors such as BBO~\cite{phinney2004big} and DECIGO~\cite{Kawamura:2011zz}.

Our results highlight the role of nonlinear dynamics in the deep eMD regime and motivate further studies that include nonlinear evolution and extended PBH mass functions. More broadly, the methodology can be applied to other realizations of matter domination, offering a framework to connect GW observations with early Universe physics.

We mainly focus on the adiabatic initial condition throughout this paper. 
However, PBHs themselves may serve as isocurvature perturbations. In this case, the existence of PBHs can produce relative entropy perturbations and source the curvature perturbation, leading to the enhancement of the power spectrum. Our result just shows that it is hard to enhance induced GWs to an observable level without enhancing the power spectrum of the curvature perturbation.

\emph{Acknowledgements.}
This work is supported in part by the National Key Research and Development Program of China Grant 
No. 2020YFC2201501, in part by the National Natural Science Foundation of China under Grant 
No. 12475067 and No. 12235019.

\bibliographystyle{apsrev4-2}
\bibliography{pbhinduce}

\begin{thebibliography}{60}%
\makeatletter
\providecommand \@ifxundefined [1]{%
 \@ifx{#1\undefined}
}%
\providecommand \@ifnum [1]{%
 \ifnum #1\expandafter \@firstoftwo
 \else \expandafter \@secondoftwo
 \fi
}%
\providecommand \@ifx [1]{%
 \ifx #1\expandafter \@firstoftwo
 \else \expandafter \@secondoftwo
 \fi
}%
\providecommand \natexlab [1]{#1}%
\providecommand \enquote  [1]{``#1''}%
\providecommand \bibnamefont  [1]{#1}%
\providecommand \bibfnamefont [1]{#1}%
\providecommand \citenamefont [1]{#1}%
\providecommand \href@noop [0]{\@secondoftwo}%
\providecommand \href [0]{\begingroup \@sanitize@url \@href}%
\providecommand \@href[1]{\@@startlink{#1}\@@href}%
\providecommand \@@href[1]{\endgroup#1\@@endlink}%
\providecommand \@sanitize@url [0]{\catcode `\\12\catcode `\$12\catcode `\&12\catcode `\#12\catcode `\^12\catcode `\_12\catcode `\%12\relax}%
\providecommand \@@startlink[1]{}%
\providecommand \@@endlink[0]{}%
\providecommand \url  [0]{\begingroup\@sanitize@url \@url }%
\providecommand \@url [1]{\endgroup\@href {#1}{\urlprefix }}%
\providecommand \urlprefix  [0]{URL }%
\providecommand \Eprint [0]{\href }%
\providecommand \doibase [0]{https://doi.org/}%
\providecommand \selectlanguage [0]{\@gobble}%
\providecommand \bibinfo  [0]{\@secondoftwo}%
\providecommand \bibfield  [0]{\@secondoftwo}%
\providecommand \translation [1]{[#1]}%
\providecommand \BibitemOpen [0]{}%
\providecommand \bibitemStop [0]{}%
\providecommand \bibitemNoStop [0]{.\EOS\space}%
\providecommand \EOS [0]{\spacefactor3000\relax}%
\providecommand \BibitemShut  [1]{\csname bibitem#1\endcsname}%
\let\auto@bib@innerbib\@empty
\bibitem [{\citenamefont {Guth}(1981)}]{Guth:1980zm}%
  \BibitemOpen
  \bibfield  {author} {\bibinfo {author} {\bibfnamefont {A.~H.}\ \bibnamefont {Guth}},\ }\href {https://doi.org/10.1103/PhysRevD.23.347} {\bibfield  {journal} {\bibinfo  {journal} {Phys. Rev. D}\ }\textbf {\bibinfo {volume} {23}},\ \bibinfo {pages} {347} (\bibinfo {year} {1981})}\BibitemShut {NoStop}%
\bibitem [{\citenamefont {Starobinsky}(1980)}]{Starobinsky:1980te}%
  \BibitemOpen
  \bibfield  {author} {\bibinfo {author} {\bibfnamefont {A.~A.}\ \bibnamefont {Starobinsky}},\ }\href {https://doi.org/10.1016/0370-2693(80)90670-X} {\bibfield  {journal} {\bibinfo  {journal} {Phys. Lett. B}\ }\textbf {\bibinfo {volume} {91}},\ \bibinfo {pages} {99} (\bibinfo {year} {1980})}\BibitemShut {NoStop}%
\bibitem [{\citenamefont {Mukhanov}\ and\ \citenamefont {Chibisov}(1981)}]{Mukhanov:1981xt}%
  \BibitemOpen
  \bibfield  {author} {\bibinfo {author} {\bibfnamefont {V.~F.}\ \bibnamefont {Mukhanov}}\ and\ \bibinfo {author} {\bibfnamefont {G.~V.}\ \bibnamefont {Chibisov}},\ }\href@noop {} {\bibfield  {journal} {\bibinfo  {journal} {JETP Lett.}\ }\textbf {\bibinfo {volume} {33}},\ \bibinfo {pages} {532} (\bibinfo {year} {1981})}\BibitemShut {NoStop}%
\bibitem [{\citenamefont {Guzzetti}\ \emph {et~al.}(2016)\citenamefont {Guzzetti}, \citenamefont {Bartolo}, \citenamefont {Liguori},\ and\ \citenamefont {Matarrese}}]{guzzetti_gravitational_2016}%
  \BibitemOpen
  \bibfield  {author} {\bibinfo {author} {\bibfnamefont {M.~C.}\ \bibnamefont {Guzzetti}}, \bibinfo {author} {\bibfnamefont {N.}~\bibnamefont {Bartolo}}, \bibinfo {author} {\bibfnamefont {M.}~\bibnamefont {Liguori}},\ and\ \bibinfo {author} {\bibfnamefont {S.}~\bibnamefont {Matarrese}},\ }\href {https://doi.org/10.1393/ncr/i2016-10127-1} {\bibfield  {journal} {\bibinfo  {journal} {La Rivista del Nuovo Cimento}\ }\textbf {\bibinfo {volume} {39}},\ \bibinfo {pages} {399} (\bibinfo {year} {2016})},\ \bibinfo {note} {arXiv: 1605.01615}\BibitemShut {NoStop}%
\bibitem [{\citenamefont {Riotto}(2017)}]{riotto_inflation_2017}%
  \BibitemOpen
  \bibfield  {author} {\bibinfo {author} {\bibfnamefont {A.}~\bibnamefont {Riotto}},\ }\href {http://arxiv.org/abs/hep-ph/0210162} {\bibfield  {journal} {\bibinfo  {journal} {arXiv:hep-ph/0210162}\ } (\bibinfo {year} {2017})},\ \bibinfo {note} {arXiv: hep-ph/0210162}\BibitemShut {NoStop}%
\bibitem [{\citenamefont {Witten}(1984)}]{Witten:1984rs}%
  \BibitemOpen
  \bibfield  {author} {\bibinfo {author} {\bibfnamefont {E.}~\bibnamefont {Witten}},\ }\href {https://doi.org/10.1103/PhysRevD.30.272} {\bibfield  {journal} {\bibinfo  {journal} {Phys. Rev. D}\ }\textbf {\bibinfo {volume} {30}},\ \bibinfo {pages} {272} (\bibinfo {year} {1984})}\BibitemShut {NoStop}%
\bibitem [{\citenamefont {Kosowsky}\ \emph {et~al.}(1992)\citenamefont {Kosowsky}, \citenamefont {Turner},\ and\ \citenamefont {Watkins}}]{Kosowsky:1991ua}%
  \BibitemOpen
  \bibfield  {author} {\bibinfo {author} {\bibfnamefont {A.}~\bibnamefont {Kosowsky}}, \bibinfo {author} {\bibfnamefont {M.~S.}\ \bibnamefont {Turner}},\ and\ \bibinfo {author} {\bibfnamefont {R.}~\bibnamefont {Watkins}},\ }\href {https://doi.org/10.1103/PhysRevD.45.4514} {\bibfield  {journal} {\bibinfo  {journal} {Phys. Rev. D}\ }\textbf {\bibinfo {volume} {45}},\ \bibinfo {pages} {4514} (\bibinfo {year} {1992})}\BibitemShut {NoStop}%
\bibitem [{\citenamefont {Hindmarsh}\ \emph {et~al.}(2021)\citenamefont {Hindmarsh}, \citenamefont {Lüben}, \citenamefont {Lumma},\ and\ \citenamefont {Pauly}}]{Hindmarsh_2021}%
  \BibitemOpen
  \bibfield  {author} {\bibinfo {author} {\bibfnamefont {M.}~\bibnamefont {Hindmarsh}}, \bibinfo {author} {\bibfnamefont {M.}~\bibnamefont {Lüben}}, \bibinfo {author} {\bibfnamefont {J.}~\bibnamefont {Lumma}},\ and\ \bibinfo {author} {\bibfnamefont {M.}~\bibnamefont {Pauly}},\ }\bibfield  {journal} {\bibinfo  {journal} {SciPost Physics Lecture Notes}\ }\href {https://doi.org/10.21468/scipostphyslectnotes.24} {10.21468/scipostphyslectnotes.24} (\bibinfo {year} {2021})\BibitemShut {NoStop}%
\bibitem [{\citenamefont {Mazumdar}\ and\ \citenamefont {White}(2019)}]{Mazumdar_2019}%
  \BibitemOpen
  \bibfield  {author} {\bibinfo {author} {\bibfnamefont {A.}~\bibnamefont {Mazumdar}}\ and\ \bibinfo {author} {\bibfnamefont {G.}~\bibnamefont {White}},\ }\href {https://doi.org/10.1088/1361-6633/ab1f55} {\bibfield  {journal} {\bibinfo  {journal} {Reports on Progress in Physics}\ }\textbf {\bibinfo {volume} {82}},\ \bibinfo {pages} {076901} (\bibinfo {year} {2019})}\BibitemShut {NoStop}%
\bibitem [{\citenamefont {Kofman}\ \emph {et~al.}(1997)\citenamefont {Kofman}, \citenamefont {Linde},\ and\ \citenamefont {Starobinsky}}]{Kofman_1997}%
  \BibitemOpen
  \bibfield  {author} {\bibinfo {author} {\bibfnamefont {L.}~\bibnamefont {Kofman}}, \bibinfo {author} {\bibfnamefont {A.}~\bibnamefont {Linde}},\ and\ \bibinfo {author} {\bibfnamefont {A.~A.}\ \bibnamefont {Starobinsky}},\ }\href {https://doi.org/10.1103/physrevd.56.3258} {\bibfield  {journal} {\bibinfo  {journal} {Physical Review D}\ }\textbf {\bibinfo {volume} {56}},\ \bibinfo {pages} {3258–3295} (\bibinfo {year} {1997})}\BibitemShut {NoStop}%
\bibitem [{\citenamefont {Greene}\ \emph {et~al.}(1997)\citenamefont {Greene}, \citenamefont {Kofman}, \citenamefont {Linde},\ and\ \citenamefont {Starobinsky}}]{Greene:1997fu}%
  \BibitemOpen
  \bibfield  {author} {\bibinfo {author} {\bibfnamefont {P.~B.}\ \bibnamefont {Greene}}, \bibinfo {author} {\bibfnamefont {L.}~\bibnamefont {Kofman}}, \bibinfo {author} {\bibfnamefont {A.~D.}\ \bibnamefont {Linde}},\ and\ \bibinfo {author} {\bibfnamefont {A.~A.}\ \bibnamefont {Starobinsky}},\ }\href {https://doi.org/10.1103/PhysRevD.56.6175} {\bibfield  {journal} {\bibinfo  {journal} {Phys. Rev. D}\ }\textbf {\bibinfo {volume} {56}},\ \bibinfo {pages} {6175} (\bibinfo {year} {1997})},\ \Eprint {https://arxiv.org/abs/hep-ph/9705347} {arXiv:hep-ph/9705347} \BibitemShut {NoStop}%
\bibitem [{\citenamefont {Felder}\ \emph {et~al.}(2001)\citenamefont {Felder}, \citenamefont {Kofman},\ and\ \citenamefont {Linde}}]{Felder:2001kt}%
  \BibitemOpen
  \bibfield  {author} {\bibinfo {author} {\bibfnamefont {G.~N.}\ \bibnamefont {Felder}}, \bibinfo {author} {\bibfnamefont {L.}~\bibnamefont {Kofman}},\ and\ \bibinfo {author} {\bibfnamefont {A.~D.}\ \bibnamefont {Linde}},\ }\href {https://doi.org/10.1103/PhysRevD.64.123517} {\bibfield  {journal} {\bibinfo  {journal} {Phys. Rev. D}\ }\textbf {\bibinfo {volume} {64}},\ \bibinfo {pages} {123517} (\bibinfo {year} {2001})},\ \Eprint {https://arxiv.org/abs/hep-th/0106179} {arXiv:hep-th/0106179} \BibitemShut {NoStop}%
\bibitem [{\citenamefont {Dufaux}\ \emph {et~al.}(2007)\citenamefont {Dufaux}, \citenamefont {Bergman}, \citenamefont {Felder}, \citenamefont {Kofman},\ and\ \citenamefont {Uzan}}]{Dufaux_2007}%
  \BibitemOpen
  \bibfield  {author} {\bibinfo {author} {\bibfnamefont {J.-F.}\ \bibnamefont {Dufaux}}, \bibinfo {author} {\bibfnamefont {A.}~\bibnamefont {Bergman}}, \bibinfo {author} {\bibfnamefont {G.}~\bibnamefont {Felder}}, \bibinfo {author} {\bibfnamefont {L.}~\bibnamefont {Kofman}},\ and\ \bibinfo {author} {\bibfnamefont {J.-P.}\ \bibnamefont {Uzan}},\ }\bibfield  {journal} {\bibinfo  {journal} {Physical Review D}\ }\textbf {\bibinfo {volume} {76}},\ \href {https://doi.org/10.1103/physrevd.76.123517} {10.1103/physrevd.76.123517} (\bibinfo {year} {2007})\BibitemShut {NoStop}%
\bibitem [{\citenamefont {Vilenkin}(1985)}]{VILENKIN1985263}%
  \BibitemOpen
  \bibfield  {author} {\bibinfo {author} {\bibfnamefont {A.}~\bibnamefont {Vilenkin}},\ }\href {https://doi.org/https://doi.org/10.1016/0370-1573(85)90033-X} {\bibfield  {journal} {\bibinfo  {journal} {Physics Reports}\ }\textbf {\bibinfo {volume} {121}},\ \bibinfo {pages} {263} (\bibinfo {year} {1985})}\BibitemShut {NoStop}%
\bibitem [{\citenamefont {Hindmarsh}\ and\ \citenamefont {Kibble}(1995)}]{Hindmarsh_1995}%
  \BibitemOpen
  \bibfield  {author} {\bibinfo {author} {\bibfnamefont {M.~B.}\ \bibnamefont {Hindmarsh}}\ and\ \bibinfo {author} {\bibfnamefont {T.~W.~B.}\ \bibnamefont {Kibble}},\ }\href {https://doi.org/10.1088/0034-4885/58/5/001} {\bibfield  {journal} {\bibinfo  {journal} {Reports on Progress in Physics}\ }\textbf {\bibinfo {volume} {58}},\ \bibinfo {pages} {477–562} (\bibinfo {year} {1995})}\BibitemShut {NoStop}%
\bibitem [{\citenamefont {Saikawa}(2017)}]{Saikawa_2017}%
  \BibitemOpen
  \bibfield  {author} {\bibinfo {author} {\bibfnamefont {K.}~\bibnamefont {Saikawa}},\ }\href {https://doi.org/10.3390/universe3020040} {\bibfield  {journal} {\bibinfo  {journal} {Universe}\ }\textbf {\bibinfo {volume} {3}},\ \bibinfo {pages} {40} (\bibinfo {year} {2017})}\BibitemShut {NoStop}%
\bibitem [{\citenamefont {Ananda}\ \emph {et~al.}(2007)\citenamefont {Ananda}, \citenamefont {Clarkson},\ and\ \citenamefont {Wands}}]{Ananda_2007}%
  \BibitemOpen
  \bibfield  {author} {\bibinfo {author} {\bibfnamefont {K.~N.}\ \bibnamefont {Ananda}}, \bibinfo {author} {\bibfnamefont {C.}~\bibnamefont {Clarkson}},\ and\ \bibinfo {author} {\bibfnamefont {D.}~\bibnamefont {Wands}},\ }\bibfield  {journal} {\bibinfo  {journal} {Physical Review D}\ }\textbf {\bibinfo {volume} {75}},\ \href {https://doi.org/10.1103/physrevd.75.123518} {10.1103/physrevd.75.123518} (\bibinfo {year} {2007})\BibitemShut {NoStop}%
\bibitem [{\citenamefont {Baumann}\ \emph {et~al.}(2007)\citenamefont {Baumann}, \citenamefont {Steinhardt}, \citenamefont {Takahashi},\ and\ \citenamefont {Ichiki}}]{Baumann_2007}%
  \BibitemOpen
  \bibfield  {author} {\bibinfo {author} {\bibfnamefont {D.}~\bibnamefont {Baumann}}, \bibinfo {author} {\bibfnamefont {P.}~\bibnamefont {Steinhardt}}, \bibinfo {author} {\bibfnamefont {K.}~\bibnamefont {Takahashi}},\ and\ \bibinfo {author} {\bibfnamefont {K.}~\bibnamefont {Ichiki}},\ }\bibfield  {journal} {\bibinfo  {journal} {Physical Review D}\ }\textbf {\bibinfo {volume} {76}},\ \href {https://doi.org/10.1103/physrevd.76.084019} {10.1103/physrevd.76.084019} (\bibinfo {year} {2007})\BibitemShut {NoStop}%
\bibitem [{\citenamefont {Bartolo}\ \emph {et~al.}(2007)\citenamefont {Bartolo}, \citenamefont {Matarrese}, \citenamefont {Riotto},\ and\ \citenamefont {Väihkönen}}]{Bartolo_2007}%
  \BibitemOpen
  \bibfield  {author} {\bibinfo {author} {\bibfnamefont {N.}~\bibnamefont {Bartolo}}, \bibinfo {author} {\bibfnamefont {S.}~\bibnamefont {Matarrese}}, \bibinfo {author} {\bibfnamefont {A.}~\bibnamefont {Riotto}},\ and\ \bibinfo {author} {\bibfnamefont {A.}~\bibnamefont {Väihkönen}},\ }\bibfield  {journal} {\bibinfo  {journal} {Physical Review D}\ }\textbf {\bibinfo {volume} {76}},\ \href {https://doi.org/10.1103/physrevd.76.061302} {10.1103/physrevd.76.061302} (\bibinfo {year} {2007})\BibitemShut {NoStop}%
\bibitem [{\citenamefont {Mangilli}\ \emph {et~al.}(2008)\citenamefont {Mangilli}, \citenamefont {Bartolo}, \citenamefont {Matarrese},\ and\ \citenamefont {Riotto}}]{Mangilli_2008}%
  \BibitemOpen
  \bibfield  {author} {\bibinfo {author} {\bibfnamefont {A.}~\bibnamefont {Mangilli}}, \bibinfo {author} {\bibfnamefont {N.}~\bibnamefont {Bartolo}}, \bibinfo {author} {\bibfnamefont {S.}~\bibnamefont {Matarrese}},\ and\ \bibinfo {author} {\bibfnamefont {A.}~\bibnamefont {Riotto}},\ }\bibfield  {journal} {\bibinfo  {journal} {Physical Review D}\ }\textbf {\bibinfo {volume} {78}},\ \href {https://doi.org/10.1103/physrevd.78.083517} {10.1103/physrevd.78.083517} (\bibinfo {year} {2008})\BibitemShut {NoStop}%
\bibitem [{\citenamefont {Saito}\ and\ \citenamefont {Yokoyama}(2009)}]{Saito_2009}%
  \BibitemOpen
  \bibfield  {author} {\bibinfo {author} {\bibfnamefont {R.}~\bibnamefont {Saito}}\ and\ \bibinfo {author} {\bibfnamefont {J.}~\bibnamefont {Yokoyama}},\ }\bibfield  {journal} {\bibinfo  {journal} {Physical Review Letters}\ }\textbf {\bibinfo {volume} {102}},\ \href {https://doi.org/10.1103/physrevlett.102.161101} {10.1103/physrevlett.102.161101} (\bibinfo {year} {2009})\BibitemShut {NoStop}%
\bibitem [{\citenamefont {Assadullahi}\ and\ \citenamefont {Wands}(2009)}]{assadullahi_gravitational_2009}%
  \BibitemOpen
  \bibfield  {author} {\bibinfo {author} {\bibfnamefont {H.}~\bibnamefont {Assadullahi}}\ and\ \bibinfo {author} {\bibfnamefont {D.}~\bibnamefont {Wands}},\ }\href {https://doi.org/10.1103/PhysRevD.79.083511} {\bibfield  {journal} {\bibinfo  {journal} {Physical Review D}\ }\textbf {\bibinfo {volume} {79}},\ \bibinfo {pages} {083511} (\bibinfo {year} {2009})},\ \bibinfo {note} {arXiv:0901.0989 [astro-ph, physics:gr-qc, physics:hep-ph]}\BibitemShut {NoStop}%
\bibitem [{\citenamefont {Assadullahi}\ and\ \citenamefont {Wands}(2010)}]{Assadullahi_2010}%
  \BibitemOpen
  \bibfield  {author} {\bibinfo {author} {\bibfnamefont {H.}~\bibnamefont {Assadullahi}}\ and\ \bibinfo {author} {\bibfnamefont {D.}~\bibnamefont {Wands}},\ }\bibfield  {journal} {\bibinfo  {journal} {Physical Review D}\ }\textbf {\bibinfo {volume} {81}},\ \href {https://doi.org/10.1103/physrevd.81.023527} {10.1103/physrevd.81.023527} (\bibinfo {year} {2010})\BibitemShut {NoStop}%
\bibitem [{\citenamefont {Kawasaki}\ \emph {et~al.}(2013)\citenamefont {Kawasaki}, \citenamefont {Kitajima},\ and\ \citenamefont {Yokoyama}}]{Kawasaki_2013}%
  \BibitemOpen
  \bibfield  {author} {\bibinfo {author} {\bibfnamefont {M.}~\bibnamefont {Kawasaki}}, \bibinfo {author} {\bibfnamefont {N.}~\bibnamefont {Kitajima}},\ and\ \bibinfo {author} {\bibfnamefont {S.}~\bibnamefont {Yokoyama}},\ }\href {https://doi.org/10.1088/1475-7516/2013/08/042} {\bibfield  {journal} {\bibinfo  {journal} {Journal of Cosmology and Astroparticle Physics}\ }\textbf {\bibinfo {volume} {2013}}\bibinfo  {number} { (08)},\ \bibinfo {pages} {042–042}}\BibitemShut {NoStop}%
\bibitem [{\citenamefont {Kohri}\ and\ \citenamefont {Terada}(2018)}]{Kohri_2018}%
  \BibitemOpen
\bibfield  {number} {  }\bibfield  {author} {\bibinfo {author} {\bibfnamefont {K.}~\bibnamefont {Kohri}}\ and\ \bibinfo {author} {\bibfnamefont {T.}~\bibnamefont {Terada}},\ }\bibfield  {journal} {\bibinfo  {journal} {Physical Review D}\ }\textbf {\bibinfo {volume} {97}},\ \href {https://doi.org/10.1103/physrevd.97.123532} {10.1103/physrevd.97.123532} (\bibinfo {year} {2018})\BibitemShut {NoStop}%
\bibitem [{\citenamefont {Domènech}(2020)}]{Dom_nech_2020}%
  \BibitemOpen
  \bibfield  {author} {\bibinfo {author} {\bibfnamefont {G.}~\bibnamefont {Domènech}},\ }\href {https://doi.org/10.1142/s0218271820500285} {\bibfield  {journal} {\bibinfo  {journal} {International Journal of Modern Physics D}\ }\textbf {\bibinfo {volume} {29}},\ \bibinfo {pages} {2050028} (\bibinfo {year} {2020})}\BibitemShut {NoStop}%
\bibitem [{\citenamefont {Zhang}(2015)}]{Zhang_2015}%
  \BibitemOpen
  \bibfield  {author} {\bibinfo {author} {\bibfnamefont {Y.}~\bibnamefont {Zhang}},\ }\href {https://doi.org/10.1088/1475-7516/2015/05/008} {\bibfield  {journal} {\bibinfo  {journal} {Journal of Cosmology and Astroparticle Physics}\ }\textbf {\bibinfo {volume} {2015}}\bibinfo  {number} { (05)},\ \bibinfo {pages} {008–008}}\BibitemShut {NoStop}%
\bibitem [{\citenamefont {Hook}\ \emph {et~al.}(2021)\citenamefont {Hook}, \citenamefont {Marques-Tavares},\ and\ \citenamefont {Racco}}]{hook_causal_2021}%
  \BibitemOpen
\bibfield  {number} {  }\bibfield  {author} {\bibinfo {author} {\bibfnamefont {A.}~\bibnamefont {Hook}}, \bibinfo {author} {\bibfnamefont {G.}~\bibnamefont {Marques-Tavares}},\ and\ \bibinfo {author} {\bibfnamefont {D.}~\bibnamefont {Racco}},\ }\href {https://doi.org/10.1007/JHEP02(2021)117} {\bibfield  {journal} {\bibinfo  {journal} {Journal of High Energy Physics}\ }\textbf {\bibinfo {volume} {2021}},\ \bibinfo {pages} {117} (\bibinfo {year} {2021})},\ \bibinfo {note} {arXiv:2010.03568 [astro-ph, physics:gr-qc, physics:hep-ph]}\BibitemShut {NoStop}%
\bibitem [{\citenamefont {Erickcek}\ \emph {et~al.}(2021)\citenamefont {Erickcek}, \citenamefont {Ralegankar},\ and\ \citenamefont {Shelton}}]{Erickcek_2021}%
  \BibitemOpen
  \bibfield  {author} {\bibinfo {author} {\bibfnamefont {A.~L.}\ \bibnamefont {Erickcek}}, \bibinfo {author} {\bibfnamefont {P.}~\bibnamefont {Ralegankar}},\ and\ \bibinfo {author} {\bibfnamefont {J.}~\bibnamefont {Shelton}},\ }\bibfield  {journal} {\bibinfo  {journal} {Physical Review D}\ }\textbf {\bibinfo {volume} {103}},\ \href {https://doi.org/10.1103/physrevd.103.103508} {10.1103/physrevd.103.103508} (\bibinfo {year} {2021})\BibitemShut {NoStop}%
\bibitem [{\citenamefont {Erickcek}\ \emph {et~al.}(2022)\citenamefont {Erickcek}, \citenamefont {Ralegankar},\ and\ \citenamefont {Shelton}}]{Erickcek_2022}%
  \BibitemOpen
  \bibfield  {author} {\bibinfo {author} {\bibfnamefont {A.~L.}\ \bibnamefont {Erickcek}}, \bibinfo {author} {\bibfnamefont {P.}~\bibnamefont {Ralegankar}},\ and\ \bibinfo {author} {\bibfnamefont {J.}~\bibnamefont {Shelton}},\ }\href {https://doi.org/10.1088/1475-7516/2022/01/017} {\bibfield  {journal} {\bibinfo  {journal} {Journal of Cosmology and Astroparticle Physics}\ }\textbf {\bibinfo {volume} {2022}}\bibinfo  {number} { (01)},\ \bibinfo {pages} {017}}\BibitemShut {NoStop}%
\bibitem [{\citenamefont {Nelson}\ and\ \citenamefont {Xiao}(2018)}]{Nelson_2018}%
  \BibitemOpen
\bibfield  {number} {  }\bibfield  {author} {\bibinfo {author} {\bibfnamefont {A.~E.}\ \bibnamefont {Nelson}}\ and\ \bibinfo {author} {\bibfnamefont {H.}~\bibnamefont {Xiao}},\ }\bibfield  {journal} {\bibinfo  {journal} {Physical Review D}\ }\textbf {\bibinfo {volume} {98}},\ \href {https://doi.org/10.1103/physrevd.98.063516} {10.1103/physrevd.98.063516} (\bibinfo {year} {2018})\BibitemShut {NoStop}%
\bibitem [{\citenamefont {Blinov}\ \emph {et~al.}(2020)\citenamefont {Blinov}, \citenamefont {Dolan},\ and\ \citenamefont {Draper}}]{Blinov_2020}%
  \BibitemOpen
  \bibfield  {author} {\bibinfo {author} {\bibfnamefont {N.}~\bibnamefont {Blinov}}, \bibinfo {author} {\bibfnamefont {M.~J.}\ \bibnamefont {Dolan}},\ and\ \bibinfo {author} {\bibfnamefont {P.}~\bibnamefont {Draper}},\ }\bibfield  {journal} {\bibinfo  {journal} {Physical Review D}\ }\textbf {\bibinfo {volume} {101}},\ \href {https://doi.org/10.1103/physrevd.101.035002} {10.1103/physrevd.101.035002} (\bibinfo {year} {2020})\BibitemShut {NoStop}%
\bibitem [{\citenamefont {Sasaki}\ \emph {et~al.}(2018)\citenamefont {Sasaki}, \citenamefont {Suyama}, \citenamefont {Tanaka},\ and\ \citenamefont {Yokoyama}}]{Sasaki:2018dmp}%
  \BibitemOpen
  \bibfield  {author} {\bibinfo {author} {\bibfnamefont {M.}~\bibnamefont {Sasaki}}, \bibinfo {author} {\bibfnamefont {T.}~\bibnamefont {Suyama}}, \bibinfo {author} {\bibfnamefont {T.}~\bibnamefont {Tanaka}},\ and\ \bibinfo {author} {\bibfnamefont {S.}~\bibnamefont {Yokoyama}},\ }\href {https://doi.org/10.1088/1361-6382/aaa7b4} {\bibfield  {journal} {\bibinfo  {journal} {Class. Quant. Grav.}\ }\textbf {\bibinfo {volume} {35}},\ \bibinfo {pages} {063001} (\bibinfo {year} {2018})},\ \Eprint {https://arxiv.org/abs/1801.05235} {arXiv:1801.05235 [astro-ph.CO]} \BibitemShut {NoStop}%
\bibitem [{\citenamefont {Inomata}\ \emph {et~al.}(2019{\natexlab{a}})\citenamefont {Inomata}, \citenamefont {Kohri}, \citenamefont {Nakama},\ and\ \citenamefont {Terada}}]{inomata_enhancement_2019}%
  \BibitemOpen
  \bibfield  {author} {\bibinfo {author} {\bibfnamefont {K.}~\bibnamefont {Inomata}}, \bibinfo {author} {\bibfnamefont {K.}~\bibnamefont {Kohri}}, \bibinfo {author} {\bibfnamefont {T.}~\bibnamefont {Nakama}},\ and\ \bibinfo {author} {\bibfnamefont {T.}~\bibnamefont {Terada}},\ }\href {https://doi.org/10.1103/PhysRevD.100.043532} {\bibfield  {journal} {\bibinfo  {journal} {Physical Review D}\ }\textbf {\bibinfo {volume} {100}},\ \bibinfo {pages} {043532} (\bibinfo {year} {2019}{\natexlab{a}})},\ \bibinfo {note} {arXiv:1904.12879 [astro-ph, physics:gr-qc, physics:hep-ph]}\BibitemShut {NoStop}%
\bibitem [{\citenamefont {Inomata}\ \emph {et~al.}(2019{\natexlab{b}})\citenamefont {Inomata}, \citenamefont {Kohri}, \citenamefont {Nakama},\ and\ \citenamefont {Terada}}]{inomata_gravitational_2019}%
  \BibitemOpen
  \bibfield  {author} {\bibinfo {author} {\bibfnamefont {K.}~\bibnamefont {Inomata}}, \bibinfo {author} {\bibfnamefont {K.}~\bibnamefont {Kohri}}, \bibinfo {author} {\bibfnamefont {T.}~\bibnamefont {Nakama}},\ and\ \bibinfo {author} {\bibfnamefont {T.}~\bibnamefont {Terada}},\ }\href {https://doi.org/10.1088/1475-7516/2019/10/071} {\bibfield  {journal} {\bibinfo  {journal} {Journal of Cosmology and Astroparticle Physics}\ }\textbf {\bibinfo {volume} {2019}}\bibfield  {number} {\bibinfo  {number} { (10)},\ \bibinfo {pages} {071}},\ }\bibinfo {note} {arXiv:1904.12878 [astro-ph, physics:gr-qc, physics:hep-ph]}\BibitemShut {NoStop}%
\bibitem [{\citenamefont {Carr}\ \emph {et~al.}(2021)\citenamefont {Carr}, \citenamefont {Kohri}, \citenamefont {Sendouda},\ and\ \citenamefont {Yokoyama}}]{carr_constraints_2021}%
  \BibitemOpen
  \bibfield  {author} {\bibinfo {author} {\bibfnamefont {B.}~\bibnamefont {Carr}}, \bibinfo {author} {\bibfnamefont {K.}~\bibnamefont {Kohri}}, \bibinfo {author} {\bibfnamefont {Y.}~\bibnamefont {Sendouda}},\ and\ \bibinfo {author} {\bibfnamefont {J.}~\bibnamefont {Yokoyama}},\ }\href {https://doi.org/10.1088/1361-6633/ac1e31} {\bibfield  {journal} {\bibinfo  {journal} {Reports on Progress in Physics}\ }\textbf {\bibinfo {volume} {84}},\ \bibinfo {pages} {116902} (\bibinfo {year} {2021})},\ \bibinfo {note} {arXiv:2002.12778 [astro-ph, physics:gr-qc, physics:hep-ph, physics:hep-th]}\BibitemShut {NoStop}%
\bibitem [{\citenamefont {Escrivà}\ \emph {et~al.}(2023)\citenamefont {Escrivà}, \citenamefont {Kuhnel},\ and\ \citenamefont {Tada}}]{escriva_primordial_2023}%
  \BibitemOpen
  \bibfield  {author} {\bibinfo {author} {\bibfnamefont {A.}~\bibnamefont {Escrivà}}, \bibinfo {author} {\bibfnamefont {F.}~\bibnamefont {Kuhnel}},\ and\ \bibinfo {author} {\bibfnamefont {Y.}~\bibnamefont {Tada}},\ }\href {http://arxiv.org/abs/2211.05767} {\bibinfo {title} {Primordial {Black} {Holes}}} (\bibinfo {year} {2023}),\ \bibinfo {note} {arXiv:2211.05767 [astro-ph, physics:gr-qc, physics:hep-ph, physics:hep-th]}\BibitemShut {NoStop}%
\bibitem [{\citenamefont {Escrivà}\ \emph {et~al.}(2024)\citenamefont {Escrivà}, \citenamefont {Tada},\ and\ \citenamefont {Yoo}}]{escriva_primordial_2024}%
  \BibitemOpen
  \bibfield  {author} {\bibinfo {author} {\bibfnamefont {A.}~\bibnamefont {Escrivà}}, \bibinfo {author} {\bibfnamefont {Y.}~\bibnamefont {Tada}},\ and\ \bibinfo {author} {\bibfnamefont {C.-M.}\ \bibnamefont {Yoo}},\ }\href {http://arxiv.org/abs/2311.17760} {\bibinfo {title} {Primordial {Black} {Holes} and {Induced} {Gravitational} {Waves} from a {Smooth} {Crossover} beyond {Standard} {Model}}} (\bibinfo {year} {2024}),\ \bibinfo {note} {arXiv:2311.17760}\BibitemShut {NoStop}%
\bibitem [{\citenamefont {Domènech}\ and\ \citenamefont {Pi}(2022)}]{Dom_nech_2022}%
  \BibitemOpen
  \bibfield  {author} {\bibinfo {author} {\bibfnamefont {G.}~\bibnamefont {Domènech}}\ and\ \bibinfo {author} {\bibfnamefont {S.}~\bibnamefont {Pi}},\ }\bibfield  {journal} {\bibinfo  {journal} {Science China Physics, Mechanics \& Astronomy}\ }\textbf {\bibinfo {volume} {65}},\ \href {https://doi.org/10.1007/s11433-021-1839-6} {10.1007/s11433-021-1839-6} (\bibinfo {year} {2022})\BibitemShut {NoStop}%
\bibitem [{\citenamefont {Jedamzik}\ \emph {et~al.}(2010)\citenamefont {Jedamzik}, \citenamefont {Lemoine},\ and\ \citenamefont {Martin}}]{Jedamzik_2010}%
  \BibitemOpen
  \bibfield  {author} {\bibinfo {author} {\bibfnamefont {K.}~\bibnamefont {Jedamzik}}, \bibinfo {author} {\bibfnamefont {M.}~\bibnamefont {Lemoine}},\ and\ \bibinfo {author} {\bibfnamefont {J.}~\bibnamefont {Martin}},\ }\href {https://doi.org/10.1088/1475-7516/2010/09/034} {\bibfield  {journal} {\bibinfo  {journal} {Journal of Cosmology and Astroparticle Physics}\ }\textbf {\bibinfo {volume} {2010}}\bibinfo  {number} { (09)},\ \bibinfo {pages} {034–034}}\BibitemShut {NoStop}%
\bibitem [{\citenamefont {Erickcek}\ and\ \citenamefont {Sigurdson}(2011)}]{Erickcek_2011}%
  \BibitemOpen
\bibfield  {number} {  }\bibfield  {author} {\bibinfo {author} {\bibfnamefont {A.~L.}\ \bibnamefont {Erickcek}}\ and\ \bibinfo {author} {\bibfnamefont {K.}~\bibnamefont {Sigurdson}},\ }\bibfield  {journal} {\bibinfo  {journal} {Physical Review D}\ }\textbf {\bibinfo {volume} {84}},\ \href {https://doi.org/10.1103/physrevd.84.083503} {10.1103/physrevd.84.083503} (\bibinfo {year} {2011})\BibitemShut {NoStop}%
\bibitem [{\citenamefont {Fan}\ \emph {et~al.}(2014)\citenamefont {Fan}, \citenamefont {Özsoy},\ and\ \citenamefont {Watson}}]{Fan_2014}%
  \BibitemOpen
  \bibfield  {author} {\bibinfo {author} {\bibfnamefont {J.}~\bibnamefont {Fan}}, \bibinfo {author} {\bibfnamefont {O.}~\bibnamefont {Özsoy}},\ and\ \bibinfo {author} {\bibfnamefont {S.}~\bibnamefont {Watson}},\ }\bibfield  {journal} {\bibinfo  {journal} {Physical Review D}\ }\textbf {\bibinfo {volume} {90}},\ \href {https://doi.org/10.1103/physrevd.90.043536} {10.1103/physrevd.90.043536} (\bibinfo {year} {2014})\BibitemShut {NoStop}%
\bibitem [{\citenamefont {Kusenko}\ and\ \citenamefont {Shaposhnikov}(1998)}]{kusenko_supersymmetric_1998}%
  \BibitemOpen
  \bibfield  {author} {\bibinfo {author} {\bibfnamefont {A.}~\bibnamefont {Kusenko}}\ and\ \bibinfo {author} {\bibfnamefont {M.}~\bibnamefont {Shaposhnikov}},\ }\href {https://doi.org/10.1016/S0370-2693(97)01375-0} {\bibfield  {journal} {\bibinfo  {journal} {Physics Letters B}\ }\textbf {\bibinfo {volume} {418}},\ \bibinfo {pages} {46} (\bibinfo {year} {1998})},\ \bibinfo {note} {arXiv:hep-ph/9709492}\BibitemShut {NoStop}%
\bibitem [{\citenamefont {Kasuya}\ \emph {et~al.}(2023)\citenamefont {Kasuya}, \citenamefont {Kawasaki},\ and\ \citenamefont {Murai}}]{Kasuya_2023}%
  \BibitemOpen
  \bibfield  {author} {\bibinfo {author} {\bibfnamefont {S.}~\bibnamefont {Kasuya}}, \bibinfo {author} {\bibfnamefont {M.}~\bibnamefont {Kawasaki}},\ and\ \bibinfo {author} {\bibfnamefont {K.}~\bibnamefont {Murai}},\ }\href {https://doi.org/10.1088/1475-7516/2023/05/053} {\bibfield  {journal} {\bibinfo  {journal} {Journal of Cosmology and Astroparticle Physics}\ }\textbf {\bibinfo {volume} {2023}}\bibinfo  {number} { (05)},\ \bibinfo {pages} {053}}\BibitemShut {NoStop}%
\bibitem [{\citenamefont {Inomata}\ \emph {et~al.}(2020)\citenamefont {Inomata}, \citenamefont {Kawasaki}, \citenamefont {Mukaida}, \citenamefont {Terada},\ and\ \citenamefont {Yanagida}}]{inomata_gravitational_2020}%
  \BibitemOpen
\bibfield  {number} {  }\bibfield  {author} {\bibinfo {author} {\bibfnamefont {K.}~\bibnamefont {Inomata}}, \bibinfo {author} {\bibfnamefont {M.}~\bibnamefont {Kawasaki}}, \bibinfo {author} {\bibfnamefont {K.}~\bibnamefont {Mukaida}}, \bibinfo {author} {\bibfnamefont {T.}~\bibnamefont {Terada}},\ and\ \bibinfo {author} {\bibfnamefont {T.~T.}\ \bibnamefont {Yanagida}},\ }\href {https://doi.org/10.1103/PhysRevD.101.123533} {\bibfield  {journal} {\bibinfo  {journal} {Physical Review D}\ }\textbf {\bibinfo {volume} {101}},\ \bibinfo {pages} {123533} (\bibinfo {year} {2020})},\ \bibinfo {note} {arXiv:2003.10455 [astro-ph, physics:gr-qc, physics:hep-ph]}\BibitemShut {NoStop}%
\bibitem [{\citenamefont {Pearce}\ \emph {et~al.}(2024)\citenamefont {Pearce}, \citenamefont {Pearce}, \citenamefont {White},\ and\ \citenamefont {Balázs}}]{pearce_gravitational_2024}%
  \BibitemOpen
  \bibfield  {author} {\bibinfo {author} {\bibfnamefont {M.}~\bibnamefont {Pearce}}, \bibinfo {author} {\bibfnamefont {L.}~\bibnamefont {Pearce}}, \bibinfo {author} {\bibfnamefont {G.}~\bibnamefont {White}},\ and\ \bibinfo {author} {\bibfnamefont {C.}~\bibnamefont {Balázs}},\ }\href {https://doi.org/10.48550/arXiv.2311.12340} {\bibinfo {title} {Gravitational {Wave} {Signals} {From} {Early} {Matter} {Domination}: {Interpolating} {Between} {Fast} and {Slow} {Transitions}}} (\bibinfo {year} {2024}),\ \bibinfo {note} {arXiv:2311.12340 [astro-ph]}\BibitemShut {NoStop}%
\bibitem [{\citenamefont {Kumar}\ \emph {et~al.}(2024)\citenamefont {Kumar}, \citenamefont {Tai},\ and\ \citenamefont {Wang}}]{kumar_towards_2024}%
  \BibitemOpen
  \bibfield  {author} {\bibinfo {author} {\bibfnamefont {S.}~\bibnamefont {Kumar}}, \bibinfo {author} {\bibfnamefont {H.}~\bibnamefont {Tai}},\ and\ \bibinfo {author} {\bibfnamefont {L.-T.}\ \bibnamefont {Wang}},\ }\href {http://arxiv.org/abs/2410.17291} {\bibinfo {title} {Towards a {Complete} {Treatment} of {Scalar}-induced {Gravitational} {Waves} with {Early} {Matter} {Domination}}} (\bibinfo {year} {2024}),\ \bibinfo {note} {arXiv:2410.17291}\BibitemShut {NoStop}%
\bibitem [{\citenamefont {Pearce}\ \emph {et~al.}(2025)\citenamefont {Pearce}, \citenamefont {Pearce}, \citenamefont {White},\ and\ \citenamefont {Balázs}}]{pearce_using_2025}%
  \BibitemOpen
  \bibfield  {author} {\bibinfo {author} {\bibfnamefont {M.}~\bibnamefont {Pearce}}, \bibinfo {author} {\bibfnamefont {L.}~\bibnamefont {Pearce}}, \bibinfo {author} {\bibfnamefont {G.}~\bibnamefont {White}},\ and\ \bibinfo {author} {\bibfnamefont {C.}~\bibnamefont {Balázs}},\ }\href {https://doi.org/10.48550/arXiv.2503.03101} {\bibinfo {title} {Using {Gravitational} {Wave} {Signals} to {Disentangle} {Early} {Matter} {Dominated} {Epochs}}} (\bibinfo {year} {2025}),\ \bibinfo {note} {arXiv:2503.03101 [astro-ph]}\BibitemShut {NoStop}%
\bibitem [{\citenamefont {Ballesteros}\ and\ \citenamefont {Taoso}(2018)}]{Ballesteros:2017fsr}%
  \BibitemOpen
  \bibfield  {author} {\bibinfo {author} {\bibfnamefont {G.}~\bibnamefont {Ballesteros}}\ and\ \bibinfo {author} {\bibfnamefont {M.}~\bibnamefont {Taoso}},\ }\href {https://doi.org/10.1103/PhysRevD.97.023501} {\bibfield  {journal} {\bibinfo  {journal} {Phys. Rev. D}\ }\textbf {\bibinfo {volume} {97}},\ \bibinfo {pages} {023501} (\bibinfo {year} {2018})},\ \Eprint {https://arxiv.org/abs/1709.05565} {arXiv:1709.05565 [hep-ph]} \BibitemShut {NoStop}%
\bibitem [{\citenamefont {Martin}\ \emph {et~al.}(2013)\citenamefont {Martin}, \citenamefont {Motohashi},\ and\ \citenamefont {Suyama}}]{Martin:2012pe}%
  \BibitemOpen
  \bibfield  {author} {\bibinfo {author} {\bibfnamefont {J.}~\bibnamefont {Martin}}, \bibinfo {author} {\bibfnamefont {H.}~\bibnamefont {Motohashi}},\ and\ \bibinfo {author} {\bibfnamefont {T.}~\bibnamefont {Suyama}},\ }\href {https://doi.org/10.1103/PhysRevD.87.023514} {\bibfield  {journal} {\bibinfo  {journal} {Phys. Rev. D}\ }\textbf {\bibinfo {volume} {87}},\ \bibinfo {pages} {023514} (\bibinfo {year} {2013})},\ \Eprint {https://arxiv.org/abs/1211.0083} {arXiv:1211.0083 [astro-ph.CO]} \BibitemShut {NoStop}%
\bibitem [{\citenamefont {Carr}\ and\ \citenamefont {Hawking}(1974)}]{Carr:1974nx}%
  \BibitemOpen
  \bibfield  {author} {\bibinfo {author} {\bibfnamefont {B.~J.}\ \bibnamefont {Carr}}\ and\ \bibinfo {author} {\bibfnamefont {S.~W.}\ \bibnamefont {Hawking}},\ }\href {https://doi.org/10.1093/mnras/168.2.399} {\bibfield  {journal} {\bibinfo  {journal} {Mon. Not. Roy. Astron. Soc.}\ }\textbf {\bibinfo {volume} {168}},\ \bibinfo {pages} {399} (\bibinfo {year} {1974})}\BibitemShut {NoStop}%
\bibitem [{\citenamefont {{Carr}}(1975)}]{1975ApJ...201....1C}%
  \BibitemOpen
  \bibfield  {author} {\bibinfo {author} {\bibfnamefont {B.~J.}\ \bibnamefont {{Carr}}},\ }\href {https://doi.org/10.1086/153853} {\bibfield  {journal} {\bibinfo  {journal} {\apj}\ }\textbf {\bibinfo {volume} {201}},\ \bibinfo {pages} {1} (\bibinfo {year} {1975})}\BibitemShut {NoStop}%
\bibitem [{\citenamefont {Dolgov}\ \emph {et~al.}(2000)\citenamefont {Dolgov}, \citenamefont {Naselsky},\ and\ \citenamefont {Novikov}}]{dolgov2000gravitationalwavesbaryogenesisdark}%
  \BibitemOpen
  \bibfield  {author} {\bibinfo {author} {\bibfnamefont {A.~D.}\ \bibnamefont {Dolgov}}, \bibinfo {author} {\bibfnamefont {P.~D.}\ \bibnamefont {Naselsky}},\ and\ \bibinfo {author} {\bibfnamefont {I.~D.}\ \bibnamefont {Novikov}},\ }\href {https://arxiv.org/abs/astro-ph/0009407} {\bibinfo {title} {Gravitational waves, baryogenesis, and dark matter from primordial black holes}} (\bibinfo {year} {2000}),\ \Eprint {https://arxiv.org/abs/astro-ph/0009407} {arXiv:astro-ph/0009407 [astro-ph]} \BibitemShut {NoStop}%
\bibitem [{\citenamefont {Aghanim}\ and\ \citenamefont {Akrami}(2020)}]{plank2020}%
  \BibitemOpen
  \bibfield  {author} {\bibinfo {author} {\bibfnamefont {N.}~\bibnamefont {Aghanim}}\ and\ \bibinfo {author} {\bibfnamefont {Y.}~\bibnamefont {Akrami}},\ }\href {https://doi.org/10.1051/0004-6361/201833910} {\bibfield  {journal} {\bibinfo  {journal} {Astronomy amp; Astrophysics}\ }\textbf {\bibinfo {volume} {641}},\ \bibinfo {pages} {A6} (\bibinfo {year} {2020})}\BibitemShut {NoStop}%
\bibitem [{\citenamefont {Baumann}(2022)}]{Baumann:2022mni}%
  \BibitemOpen
  \bibfield  {author} {\bibinfo {author} {\bibfnamefont {D.}~\bibnamefont {Baumann}},\ }\href {https://doi.org/10.1017/9781108937092} {\emph {\bibinfo {title} {{Cosmology}}}}\ (\bibinfo  {publisher} {Cambridge University Press},\ \bibinfo {year} {2022})\BibitemShut {NoStop}%
\bibitem [{\citenamefont {Ma}\ and\ \citenamefont {Bertschinger}(1995)}]{ma_cosmological_1995}%
  \BibitemOpen
  \bibfield  {author} {\bibinfo {author} {\bibfnamefont {C.-P.}\ \bibnamefont {Ma}}\ and\ \bibinfo {author} {\bibfnamefont {E.}~\bibnamefont {Bertschinger}},\ }\href {https://doi.org/10.1086/176550} {\bibfield  {journal} {\bibinfo  {journal} {The Astrophysical Journal}\ }\textbf {\bibinfo {volume} {455}},\ \bibinfo {pages} {7} (\bibinfo {year} {1995})},\ \bibinfo {note} {arXiv:astro-ph/9506072}\BibitemShut {NoStop}%
\bibitem [{\citenamefont {Abe}\ \emph {et~al.}(2021)\citenamefont {Abe}, \citenamefont {Tada},\ and\ \citenamefont {Ueda}}]{abe_induced_2021}%
  \BibitemOpen
  \bibfield  {author} {\bibinfo {author} {\bibfnamefont {K.~T.}\ \bibnamefont {Abe}}, \bibinfo {author} {\bibfnamefont {Y.}~\bibnamefont {Tada}},\ and\ \bibinfo {author} {\bibfnamefont {I.}~\bibnamefont {Ueda}},\ }\href {https://doi.org/10.1088/1475-7516/2021/06/048} {\bibfield  {journal} {\bibinfo  {journal} {Journal of Cosmology and Astroparticle Physics}\ }\textbf {\bibinfo {volume} {2021}}\bibfield  {number} {\bibinfo  {number} { (06)},\ \bibinfo {pages} {048}},\ }\bibinfo {note} {arXiv:2010.06193 [astro-ph, physics:hep-ph]}\BibitemShut {NoStop}%
\bibitem [{\citenamefont {Abe}\ and\ \citenamefont {Tada}(2023)}]{abe_translating_2023}%
  \BibitemOpen
  \bibfield  {author} {\bibinfo {author} {\bibfnamefont {K.~T.}\ \bibnamefont {Abe}}\ and\ \bibinfo {author} {\bibfnamefont {Y.}~\bibnamefont {Tada}},\ }\href {https://doi.org/10.1103/PhysRevD.108.L101304} {\bibfield  {journal} {\bibinfo  {journal} {Physical Review D}\ }\textbf {\bibinfo {volume} {108}},\ \bibinfo {pages} {L101304} (\bibinfo {year} {2023})},\ \bibinfo {note} {arXiv:2307.01653 [astro-ph]}\BibitemShut {NoStop}%
\bibitem [{\citenamefont {Phinney}\ \emph {et~al.}(2004)\citenamefont {Phinney}, \citenamefont {Bender}, \citenamefont {Buchman}, \citenamefont {Byer}, \citenamefont {Cornish}, \citenamefont {Fritschel}, \citenamefont {Folkner}, \citenamefont {Merkowitz}, \citenamefont {Danzmann}, \citenamefont {DiFiore} \emph {et~al.}}]{phinney2004big}%
  \BibitemOpen
  \bibfield  {author} {\bibinfo {author} {\bibfnamefont {S.}~\bibnamefont {Phinney}}, \bibinfo {author} {\bibfnamefont {P.}~\bibnamefont {Bender}}, \bibinfo {author} {\bibfnamefont {R.}~\bibnamefont {Buchman}}, \bibinfo {author} {\bibfnamefont {R.}~\bibnamefont {Byer}}, \bibinfo {author} {\bibfnamefont {N.}~\bibnamefont {Cornish}}, \bibinfo {author} {\bibfnamefont {P.}~\bibnamefont {Fritschel}}, \bibinfo {author} {\bibfnamefont {W.}~\bibnamefont {Folkner}}, \bibinfo {author} {\bibfnamefont {S.}~\bibnamefont {Merkowitz}}, \bibinfo {author} {\bibfnamefont {K.}~\bibnamefont {Danzmann}}, \bibinfo {author} {\bibfnamefont {L.}~\bibnamefont {DiFiore}}, \emph {et~al.},\ }\href@noop {} {\bibfield  {journal} {\bibinfo  {journal} {NASA Mission Concept Study}\ } (\bibinfo {year} {2004})}\BibitemShut {NoStop}%
\bibitem [{\citenamefont {Kawamura}\ \emph {et~al.}(2011)\citenamefont {Kawamura} \emph {et~al.}}]{Kawamura:2011zz}%
  \BibitemOpen
  \bibfield  {author} {\bibinfo {author} {\bibfnamefont {S.}~\bibnamefont {Kawamura}} \emph {et~al.},\ }\href {https://doi.org/10.1088/0264-9381/28/9/094011} {\bibfield  {journal} {\bibinfo  {journal} {Class. Quant. Grav.}\ }\textbf {\bibinfo {volume} {28}},\ \bibinfo {pages} {094011} (\bibinfo {year} {2011})}\BibitemShut {NoStop}%
\end{thebibliography}%

\end{document}